\documentclass[graybox]{svmult}

\usepackage{amssymb}
\usepackage{latexsym}
\usepackage{amsmath}
\usepackage{color}

\usepackage{mathptmx}       
\usepackage{helvet}         
\usepackage{courier}        
\usepackage{type1cm}        
\usepackage{makeidx}         
\usepackage{graphicx}        
\usepackage{multicol}        
\usepackage[bottom]{footmisc}

\newcommand{\be}{\begin{eqnarray}}
\newcommand{\ee}{\end{eqnarray}}

\makeindex             

\begin{document}

\title*{Phase transitions of regular Schwarzschild-Anti-deSitter black holes}
\author{Antonia Micol Frassino 
} 
\institute{Antonia Micol Frassino (Speaker) 
\at Frankfurt Institute for Advanced Studies (FIAS), Ruth-Moufang-Stra\ss e 1, 60438 Frankfurt am Main, Germany \&  
Johann Wolfgang Goethe-Universit\"{a}t,  Frankfurt am Main, Germany,
 \email{frassino@fias.uni-frankfurt.de} 
}
%
%
\maketitle
\abstract{We study a solution of the Einstein's equations generated by a
self-gravitating, anisotropic, static, non-singular matter fluid.  The resulting Schwarzschild like solution is regular and accounts for smearing effects of noncommutative fluctuations of the geometry. We call this solution regular Schwarzschild spacetime. In the presence of an Anti-deSitter cosmological term, the regularized metric offers an extension of the Hawking-Page transition into a van der Waals-like phase diagram. Specifically the regular Schwarzschild-Anti-deSitter geometry undergoes a first order small/large black hole transition similar to the liquid/gas transition of a real fluid. In the present analysis we have considered the cosmological constant as a dynamical quantity and its variation is included in the first law of black hole thermodynamics. }
\section{Regular Schwarzschild-anti-deSitter spacetime}
The regular Schwarzschild anti-deSitter (AdS) metric is a static, spherically symmetric solution of the Einstein's equations with negative cosmological constant $\Lambda= -3/b^2$ and a Gaussian matter  source \cite{Nicolini:2005vd,Nicolini:2008aj,Nicolini:2011dp,Smailagic:2012cu}. To obtain this metric we replace  the vacuum
with a Gaussian distribution having variance equivalent to the parameter $\sqrt\theta$
\be
\rho\left(r\right)\equiv\frac{M}{\left(4\pi\theta\right)^{3/2}}e^{-r^{2}/4\theta}
\ .
\label{eq:rho}
\ee
 This type of matter distribution emulates non-commutativity of space-time through the parameter $\theta$ that corresponds to the area of the
elementary quantum cell, accounting for a natural ultraviolet spacetime cut-off  (see Ref. \cite{Nicolini:2008aj} and references therein).
The resulting energy momentum tensor describes an anisotropic fluid, whose components, fixed by 
 $\nabla_{\mu}T^{\mu\nu}=0$ and the condition $g_{00}=-g^{-1}_{rr}$, read
\be
T_{0}^{0}=T_{r}^{r}=-\rho\left(r\right)\quad\quad T_{\phi}^{\phi}=T_{\theta}^{\theta}=-\rho\left(r\right)-\frac{r}{2}\frac{\partial\rho\left(r\right)}{\partial r}
\ .
\label{eq:components}
\ee
The spherically symmetric solution of the Einstein's equations with this energy momentum tensor and the cosmological constant $\Lambda$ is given by the line element
\be
ds^{2}=-V\left(r\right)dt^{2}+\frac{dr^{2}}{V\left(r\right)}+r^{2}d\Omega^{2}\label{eq:metric}
\ee
where $d\Omega^{2}=d\vartheta^2+\sin^2\vartheta d\varphi^2$  and 
\be
V\left(r\right)=1+\frac{r^{2}}{b^{2}}-\frac{\omega M}{r}\gamma\left(\frac{3}{2},\frac{r^{2}}{4\theta}\right)
\ .
 \label{eq:goo}
\ee
Here $\omega=2\textrm{G}_{\mathrm{N}}/\Gamma\left(3/2\right)$, $\textrm{G}_{N}$ is the four-dimensional Newton's constant and $b$ is the curvature radius of the AdS space. The  function $\gamma\left(\frac{3}{2},\frac{r^{2}}{4\theta}\right)$ is the incomplete gamma function defined as $
\gamma\left(n,x\right) \equiv \displaystyle \intop_{0}^{x}dt\, t^{n-1}e^{-t}
\ .
$
The line element \eqref{eq:metric} has an event horizon at $r = r_{+}$, where $r_{+}$ is solution of the horizon equation $V \left(r\right) = 0$. The event horizon radius coincides with the Schwarzschild radius in the limit $\sqrt{\theta} / r_{+}  \rightarrow 0$. The metric \eqref{eq:metric} admits an inner horizon $r_-<r_+$, that coalesces with $r_+$ in the extremal black hole configuration at $r_0=r_+=r_-$. Such a degenerate horizon occurs even without charge or angular momentum. 
\vspace{-0.5cm}	
\section{ Thermodynamics and Equation of state}
The temperature associated to the event horizon $r_+$ can be computed through the formula $ T=\frac{1}{4\pi}\left. V^{\prime}\left(r\right)\right|_{r=r_+}$ and reads
\be
T & = & \frac{1}{4\pi r_{+}}\left\{ 1+\frac{r_{+}^{2}}{b^{2}}\left(3-r_{+}\frac{\gamma^{\prime}\left(r_{+}\right)}{\gamma\left(r_{+}\right)}\right)-r_{+}\frac{\gamma^{\prime}\left(r_{+}\right)}{\gamma\left(r_{+}\right)}\right\} 
\ ,
\label{eq:Temperature}
\ee
where $\gamma\left(r_+\right)\equiv\gamma\left(\frac{3}{2},\,\frac{r^2_+}{4\theta}\right)$,  $\gamma'\left(r_+\right)=\frac{r_{+}^{2}}{4 \theta^{3/2}} e^{-r_{+}^{2}/4\theta}$ is its derivative  with
respect to $r_{+}$. 
In contrast to the standard Schwarzschild-anti-deSitter case, extremal solution exists with vanishing Hawking temperature \eqref{eq:Temperature}.

Recently, the idea of including the variation of the cosmological constant in the first law of black hole thermodynamics has been considered \cite{Kastor,Dolan:2010ha, Kubiznak:2012wp} with interesting consequences:
If the cosmological constant $\Lambda$ behaves like a pressure, we have that for negative cosmological constant the pressure turns to be positive \cite{Dolan:2010ha, Cvetic:2010jb}, i.e.,
\be
\frac{1}{b^{2}}=-\frac{\Lambda}{3} \equiv \frac{8\pi P}{3}
\ ,
\ee
giving rise to several effects (see for example \cite{Gunasekaran:2012dq, Altamirano:2013ane, Frassino:2014pha}). In such a case the equation of state $P\left(V,T\right)$ for the regular AdS black hole becomes
\be
P=\frac{3\gamma\left(r_{+}\right)}{\left(3\gamma\left(r_{+}\right)-r_{+}\gamma^{\prime}\left(r_{+}\right)\right)}\left\{ \frac{T}{2r_{+}}-\frac{1}{8\pi r_{+}^{2}}+\frac{\gamma^{\prime}\left(r_{+}\right)}{8\pi r_{+}\gamma\left(r_{+}\right)}\right\}. 
\label{eq:eq.state} 
\ee
Here $T$ is the Hawking temperature of the black hole, \textit{i.e.} \eqref{eq:Temperature}. 
Using the equation of state \eqref{eq:eq.state} it is possible to plot the isotherm functions in a $P-V$ diagram for a regular black hole that resembles the van der Waals pressure-volume diagram. 
%
\begin{figure}[h]
\sidecaption
\includegraphics[width=7.5cm,height=6cm]{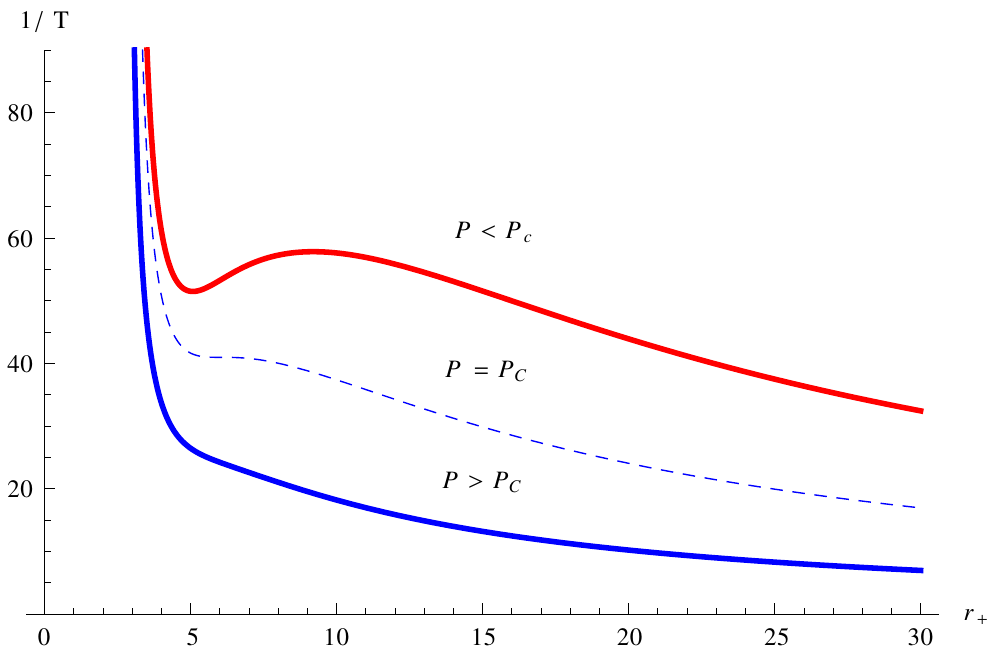}
\caption{The inverse temperature as function of $r_{+}$ (with $\theta=1$). When $P<P_{c}$, there are three branches. The middle branch is unstable, while the branch with the smaller radii and the one with bigger radii are stable. This graph reproduces the pressure-volume diagram of the van der Waals theory, provided one identifies the black hole thermodynamic variables $\beta \equiv 1/T$,  $r_{+}$ and $P$ respectively  with pressure pressure, volume  and temperature of the van der Waals gas.
 \label{fig:The-inverse-temperature} }
\end{figure}
\vspace{-0.5cm}	
\subsection{Gibbs free energy}
\begin{figure}[t]
\begin{centering}
\includegraphics[scale=0.55]{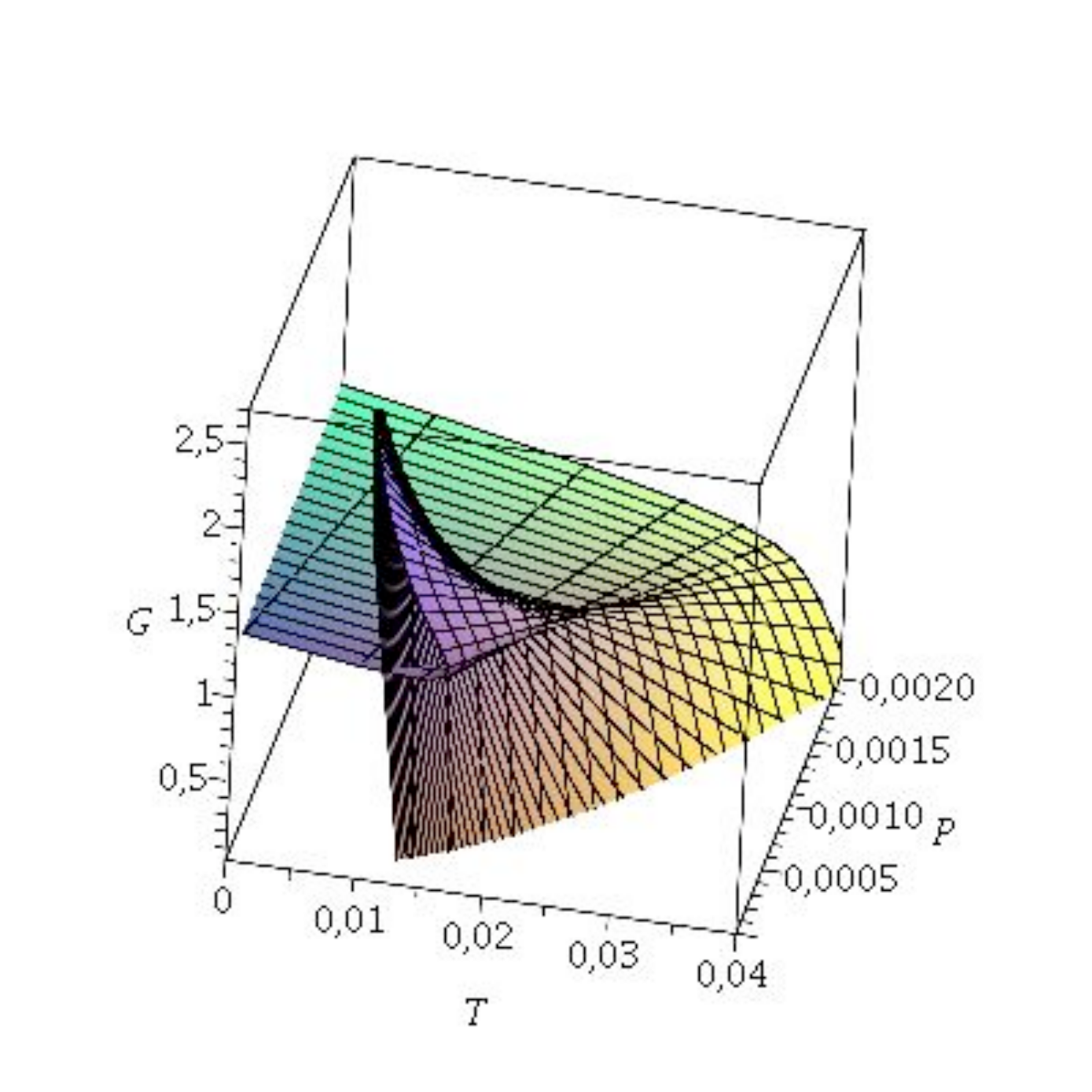} 
\par\end{centering}
\caption{
Gibbs free energy as function of the black hole pressure and temperature.
The Gibbs free energy $G$ changes its sign at a specific $T$ and $P$ (intersection of the function with the $T-P$-plane).
As in the van der Waals case, the phases are controlled by the universal `cusp', typical of the theory of discontinuous transitions \cite{Chamblin:1999tk}. The Gibbs free energy shows the ``swallowtail'' shape, a region where $G(T,P)$ is a multivalued function. This region ends in a point $(T_c,P_c)$.
In the region with $P<P_{c}$ and $T<T_{c}$ we can see a transition between small black hole/large black hole.  Note that $r_{+}$ is a function of temperature and pressure via the equation of state Eq.\eqref{eq:eq.state}. For large value of $P$ (or $T$) there is only one branch allowed.
\label{fig:wdvalls}}
\end{figure}

In order to complete the analogy between the regular black hole and a van der Waals gas,  we proceed by calculating the Gibbs free energy \cite{Dolan:2010ha,Kubiznak:2012wp}. This can be done by calculating the action of the Euclidean metric (see for example \cite{Witten:1998zw}).  Such an action provides the Gibbs free energy via $G= I/\beta$ where $\beta$ is the period of the imaginary time $\beta \equiv 1/T$. Then, the Gibbs free energy can be expressed as a function of pressure and temperature. The Hawking-Page transition \cite{HP83} for the standard Schwarzschild-AdS black hole is first order phase transition between a large black hole phase and the purely thermal AdS spacetime. Such a transition takes place when the Gibbs energy changes its sign from positive to negative. 
 In the regular black hole case and considering the cosmological constant as a pressure  we find
\be
G=
\frac{r_{+}}{12\, G_{N}}\left[3-8P\pi r_{+}^{2}+\frac{r_{+}\left(3+8P\pi r_{+}^{2}\right)\text{\ensuremath{\gamma^{\prime}\left(r_{+}\right)}}}{\gamma\left(r_{+}\right)}\right]
\label{eq:Gibbs}
\ee
and the Gibbs free energy \eqref{eq:Gibbs} exhibits a characteristic swallowtail behavior (see Fig. \ref{fig:wdvalls}). This usually corresponds to a small black hole/large black hole first-order phase transition \cite{Kubiznak:2012wp,Chamblin:1999tk}. 
By performing the classical limit for $r\gg\theta$ we get the usual result for a classical uncharged Schwarzschild-AdS black hole that is $G\left(T,P \right)= \left(1/4G_{N}\right)\left( r_{+} - \frac{8 \pi}{3}P\,r^3_{+} \right)$   \cite{Kubiznak:2012wp}. Remarkably, in the regular Schwarzschild-AdS black hole case, as in the Reissner-Nordstr\"{o}m-AdS
(RN-AdS) black hole spacetime, there is a phase transition that occurs at positive Gibbs energy. This fact is visible from the presence of the swallowtail in Fig. \ref{fig:wdvalls}. To investigate this aspect we need to study the sign of the heat capacity.
%
As underlined in \cite{Dolan:2010ha}, the specific heat related to the black hole is calculated at constant pressure 
\be
C_{p}=\left(\frac{\partial H}{\partial T}\right)_{P}=\left(\frac{\partial H}{\partial r_{+}}\right)_{P}\left(\frac{\partial r_{+}}{\partial T}\right)_{P}
\ ,
\label{eq:defCp}
\ee
where the enthalpy $H$ is identified with the black hole mass $M$ \cite{Dolan:2010ha}.
 Now one can study the phase transitions from the change of the sign of the specific heat: the stability  requires that the specific heat at fixed pressure is $C_{p}\geq0$ and the specific heat at fixed volume is $C_{v}\geq0$. 
In the case under investigation $C_{v}$  is always equal to zero because the entropy is only volume dependent. This means that the heat capacity $C_{v}$ does not diverge at the critical point and its critical exponent is $\alpha=0$. 
By studying the sign of the function $C_{p}$, we can see that for $P > P_{c}$ the quantity $C_{p}$ is always positive and the black hole is stable. In the limit $P \rightarrow P_{c}$ there is a critical value for $r_{+}$ for which $C_{p}$ diverges. For $P<P_{c}$ there are two discontinuities of the specific heat and the situation is the same as in the Reissner-Nordstr\"{o}m-AdS black holes \cite{Niu:2011tb}. Thus, in the regular Schwarzschild-AdS case for $P<P_{c}$ it seems that a different phase transition is allowed because the heat capacity changes again from positive values to negative values. For large $r_{+}$ we have the Hawking-Page behavior in which the branch with negative specific heat has lower mass and thus falls in an unstable phase, while the branch with larger mass is locally stable and corresponds to a positive specific heat. Thus, the resulting phase diagram presents a critical point at a critical cosmological constant value in Plank units and a smooth crossover thereafter.
\vspace{-0.5cm}	
\subsection{Critical Exponent}
We already determined $\alpha = 0$ in the previous section.
Now, by defining the variable $t \equiv \left(T-T_{\mathrm{c}}\right)/T_{\mathrm{c}}$, we can compute  the critical exponent of $C_{p}$ by evaluating the ratio $\ln\left(C_{p}\left(t\right)\right)/\ln\left(t\right)$ in the limit $t\to 0$. We find that the limit exists and the critical exponent is $\gamma=1$. This result implies that the heat capacity diverges near the critical point like $C_{p}\propto\left|t\right|^{-1}$. Then using the scaling relations 
\be
\alpha+2\beta+\gamma & = & 2\\
\alpha+\beta\left(1+\delta\right) & = & 2
\ee
is possible to calculate the other two exponents, i.e., $\delta$ that determines the behaviour of the isothermal compressibility of a VdW system and $\beta$ that describes the behaviour of the difference between of the volume of the gas phase and the liquid phase.
For the regular black hole, the scaling relations give $\delta=3$ and $\beta=1/2$, result that coincides with the case of charged black holes \cite{Kubiznak:2012wp}.  These critical exponents are consistent with the Ising mean field values $\left(\alpha,\beta,\gamma,\delta\right)=\left(0,1/2,1,3\right)$ allowing for an efficient mean field theory description. Since it is believed that the determination of critical exponents define universality classes , i.e., they do not depend on the details of
the physical system (exept the number of dimensions), we can say that the phase transitions in the regular Schwarzschild-AdS black holes and in the RN-AdS black holes in four-dimensional spacetime have the same nature. 
\vspace{-0.5cm}	
\section{Final remarks}
After almost hundred years since the Karl Schwarzschild's exact solution of Einstein's equation, black hole physics is nowadays at the forefront of current research in several branches of theoretical physics. 
Specific interest has been developed in the thermodynamics of charged black holes in asymptotically AdS spacetime, largely because they admit a gauge duality description via a dual thermal field theory \cite{Witten:1998zw}. \\ In recent studies  it has been shown that charged Reissner-Nordstr\"{o}m AdS black holes exhibit critical behaviour similar to a van der Waals liquid gas phase transition \cite{Kubiznak:2012wp}. This analogy become ``complete'' if the cosmological constant $\Lambda$ is considered as a dynamical quantity and its variation is included in the first law of black hole thermodynamics \cite{Gunasekaran:2012dq, Kubiznak:2012wp}. This extended phase space shows new insights with respect to the conventional phase space of a four dimensional black hole in AdS background consisting only of two variables: entropy and temperature. In this work the cosmological constant has been considered as a thermodynamical pressure and its conjugate quantity as a thermodynamical volume. The black hole equation of state \eqref{eq:eq.state} obtained by considering the regular Schwarzschild-AdS solution shows analogy with the van der Waals liquid-gas system where the parameter $\theta$ plays an analogue role of the charge. Note that a detailed description of the not-extended phase space has been presented in Ref. \cite{Nicolini:2011dp}. 
\begin{acknowledgement}
This work has been supported by the Helmholtz Research School for Quark Matter Studies (H-QM). 
The author is grateful to P. Nicolini and D. Kubiznak for having carefully read the draft and provided valuable comments.  
\end{acknowledgement}
\vspace{-0.8cm}
\end{document}